\documentclass{appolb}
\usepackage{epsfig}
\usepackage{axodraw}
\usepackage{epsfig}
\usepackage{colordvi}
\usepackage{color}
\usepackage{graphics}
\newcommand{\gtaet}{\raisebox{-0.8mm}%
{\hspace{1mm}$\stackrel{>}{\sim}$\hspace{1mm}}}

\preprint{LU TP 03-25}
\begin{document}
\title{An Intuitive Semiclassical Picture of Proton Structure at Small $x$}
\author{G\"{o}sta Gustafson
\address{Dept. of Theoretical Physics, Lund University\\
S\"{o}lvegatan 14A, S-22362 Lund, Sweden}
}

\maketitle
\begin{flushright}
  \vspace*{-9cm}
  LU TP 03-25\\
  hep-ph/0306108\\
  May 2003
  \vspace*{7cm}
\end{flushright}

Dedicated to Jan Kwieci\'{n}ski on his 65th birthday

\begin{abstract}

A semiclassical description of structure functions in DIS at small $x$ 
is presented. It gives an intuitive picture of the transition from the Double 
Leading Log approximation at large $Q^2$, to the powerlike dependence on $x$ 
in the BFKL region at limited $Q^2$. Formal derivations from perturbative 
QCD, \eg in the BFKL or the CCFM formalisms,
are technically complicated, and therefore such an intuitive picture may be 
valuable and possibly helpful in the work towards
a better understanding of DIS, and of the strong interaction in general. 

\end{abstract}
\PACS{12.38-t, 13.60.Hb}
  
\section{Introduction}

Experiments which study deep inelastic $e p$ scattering (DIS) have given 
decisive information about the basic structure of matter. Results from the 
linear accelerator at SLAC \cite{slac} gave the first evidence for a pointlike 
substructure in the proton. At the higher energies available at the HERA 
collider at DESY it has been possible to penetrate still deeper into the 
proton. DIS has the advantage over \eg $e^{+}e^-$-annihilation and hadronic 
collisions, that there are two different scales, the total 
hadronic energy $W$ and the photon virtuality $Q^2$, which can be varied 
independently. This offers a unique possibility to study the interplay between 
the perturbative and non-perturbative regimes in strong interaction.

A remarkable result at HERA is the observation that for a fixed limited
value of $Q^2$, the $\gamma^* p$ cross section grows very rapidly for large 
$W$ or small $x_{Bj} \equiv Q^2/(W^2+Q^2)$, approximately proportional to
$x_{Bj}^{-\lambda} \sim (W^2)^\lambda$ with $\lambda \sim 0.3$. Such a strong 
increase, which is significantly stronger than the corresponding growth of the 
$pp$ cross section at similar energies, was predicted by the BFKL \cite{bfkl}
evolution formalism, based on the exchange of perturbative gluon chains.
Indeed, in leading logarithmic order the BFKL equation predicts an even 
stronger increase than what is experimentally observed. 

The significance of the BFKL result 
is somewhat reduced by the facts that first the next to leading order 
correction is very large, secondly the introduction of a running coupling 
would favour chains close to the non-perturbative region, which makes the result
sensitive to a necessary soft cutoff.
Both of these effects make the perturbative expansion less reliable. Although
the experimental results 
do indicate that some hard perturbative mechanism is at work, it has
also been possible to fit the data without the BFKL mechanism, provided
the soft nonperturbative input gluon distribution grows sufficiently 
rapidly for small  $x_{Bj}$. Measurements of the total cross section,
described by the structure function $F_2$, cannot separate between the 
different mechanisms. Further experimental studies of the properties of the
final states are needed, in parallel with theoretical work to distinguish 
between different possibilities.

The derivation 
of the BFKL equation, and also of the CCFM equation \cite{ccfm} which 
interpolates between the finite-$Q^2$-BFKL and the large-$Q^2$ regimes, are 
technically quite complicated.
A semiclassical description, which can give an intuitive picture of the 
results, may therefore be valuable and possibly helpful in the work towards
a better understanding of DIS, and of the strong interaction in general.
An attempt for such a picture will be presented in the sections 2 - 5 below.

The Linked Dipole Chain model (LDC) \cite{LDC} is a reformulation and 
generalization of the CCFM model. The formalism in the LDC model has great 
similarities with the description in the semiclassical model
presented here, and thus it offers a quantitative motivation for the
qualitative results in our intuitive picture. 
A very brief description of the LDC model is presented in section 6, including
also a discussion of possible applications to hadronic collisions.

In this article I want to emphasize the similarities between QCD and
QED, and have therefore repeated wellknown results from 
electrodynamics. Large parts may therefore appear rather trivial to many 
readers, who consequently may omit these sections.

\section{Bremsstrahlung in Electrodynamics}

\subsection{Photon emission}

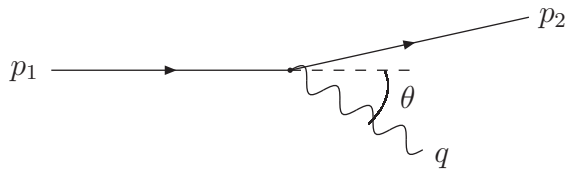
\begin{figure}
\begin{center}
\begin{picture}(200,80)(0,10) 
\ArrowLine(10,50)(100,50) \Text(0,50)[]{ { \large{$p_1$} } }
\ArrowLine(100,50)(190,70) \Text(200,70)[]{ { \large{$p_2$} } }
\DashLine(100,50)(145,50){4}
\Photon(100,50)(150,20){4}{4} \Text(158,17)[]{ { \large{$q$} } }
\qbezier(130,30)(140,40)(136,50) \Text(145,40)[]{ { \large{$\theta$} } }
\Vertex(100,50){1}
\end{picture}
\end{center}
\caption{A particle with momentum $p_1$ emits a photon with momentum $q$,
and obtains after recoil momentum $p_2=p_1-q$.}
\label{figure::bromsstralning}
\end{figure}

Let us study a charged particle with momentum $p_1$, which emits a photon 
with momentum $q$, and after recoil has obtained momentum $p_2 = p_1 - q$,
see fig.~\ref{figure::bromsstralning}. 
The interaction between a charged scalar particle and the electromagnetic 
field is described by the interaction term
\begin{equation}
\sim e j_\mu A_\mu \,\,\sim e (p_1 + p_2)_\mu \epsilon_\mu .
\label{eikonal1}
\end{equation}
This contribution depends only on the trajectory of the charged particle, and 
corresponds to what is called the ``eikonal current''. The resulting emission 
spectrum can be presented in different equivalent forms
\begin{equation}
dn \,\sim \alpha \frac{d \theta^2}{\theta^2} \frac{d q_0}{q_0}
\,\sim \alpha \frac{d q_\perp^2}{q_\perp^2} \frac{d z}{z} 
\,\sim \alpha \frac{d q_\perp^2}{q_\perp^2} d y ,
\label{soft}
\end{equation}
where $z= q_0/p_{1,0}$ is the fraction of the parent energy given to the 
photon, and $y = \frac{1}{2} \ln \frac{q_0+q_L}{q_0-q_L}$.

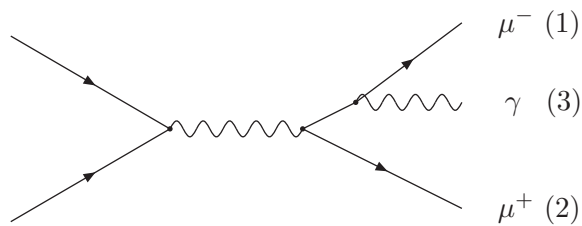
\begin{figure}
\begin{center}
\begin{picture}(200,100)(0,10)
\ArrowLine(0,85)(60,50) 
\ArrowLine(0,15)(60,50)
\Photon(60,50)(110,50){3}{5}
\Line(110,50)(130,60)
\ArrowLine(130,60)(170,90) \Text(200,90)[]{ { {$\mu^-\,\,(1)$} } }
\ArrowLine(110,50)(170,20) \Text(200,20)[]{ { {$\mu^+\,\,(2)$} } }
\Photon(130,60)(170,60){3}{4} \Text(198,60)[]{ { {$\gamma\,\,\,\,\,(3)$}}}
\Vertex(60,50){1}
\Vertex(110,50){1}
\Vertex(130,60){1}
\end{picture}
\end{center}
\caption{$e^{+}e^-$-annihilation into a $\mu^{+}\mu^-$ pair 
and a photon.}
\label{figure::mumufoton}
\end{figure}

As an example we look at $e^{+}e^-$-annihilation into a $\mu^{+}\mu^-$ pair 
and a photon, see fig.~\ref{figure::mumufoton}. If the scaled energies of the 
$\mu^{-}$, the $\mu^+$,  and the photon 
are denoted $x_1$, $x_2$, and $x_3$ respectively, the cross section is given by
\begin{equation}
dn \sim \alpha \frac{d x_1 d x_2}{(1-x_1)(1-x_2)} (x_1^2 + x_2^2) .
\label{mu+mu-}
\end{equation}
With a suitable definition of a longitudinal direction we have
\begin{equation}
q_\perp^2 = s(1-x_1)(1-x_2), \,\,\,y=\frac{1}{2} \ln(\frac{1-x_1}{1-x_2}) .
\label{kt-y-def}
\end{equation} 
With this definition the factor 
$d x_1 d x_3/(1-x_1)(1-x_3)$ in eq.~(\ref{mu+mu-}) is identical to the
expression in eq.~(\ref{soft}), $(d q_\perp^2 / q_\perp^2) d y$. 
The last factor in eq.~(\ref{mu+mu-}) is a correction factor due to the muon 
spin.

For collinear emission $q_\perp \rightarrow 0$. The squared propagator gives 
a factor $\sim 1/q_\perp^4$, but for such emissions we have 
($\mathbf{p}_1+\mathbf{p}_2) \,/\!/\,\mathbf{q} \,\bot \,\mathbf{\epsilon}$, 
which implies a suppression such that the net result is proportional to the
factor $d q_\perp^2 / q_\perp^2$ in eq.~(\ref{soft}). Thus we note that the 
vector nature of the photon is essential.

\subsection{Ordered cascades}

\emph{Sudakov form factors}

The total emission probability in eq.~(\ref{mu+mu-}) diverges, which thus gives
an infinite cross section. To order $\alpha$ this infinity is compensated 
by virtual corrections to the probability for no emission, which is a 
negative contribution. Thus in leading order a cutoff is needed in 
$q_\perp$ or in invariant masses $M_{\mu^+, \gamma}$ and $M_{\gamma, \mu^-}$. 
With such a cutoff the cross sections for zero and one emission are both
positive, and their sum, the first order result for the total cross section, 
is approximately the same as the zero's order expression. This result can be 
generalized to higher orders. In higher orders the probability to emit an 
extra photon is also compensated by a reduced probability for no emission, in 
such a way that the total cross section is (approximately) unaffected.
Thus the annihilation process can be separated into two independent phases, 
the initial production of the $\mu^+\mu^-$ pair, and the subsequent photon 
bremsstrahlung.

Soft emissions, for which the recoils can be neglected, 
are also fully uncorrelated, and the probability to emit a fixed 
number of photons in a given phase space region is given by a Poissonian 
distribution. This implies that the probability 
for no photon emission in a certain excluded region exponentiates, and is 
proportional to the ``Sudakov form factor''
\begin{equation}
S = \exp (-\int_{excl.\,\, region} d n).
\label{sudakov}
\end{equation}

This result is quite general.
For processes in which an emission does not change the total 
reaction probability, the emission probability is normally 
associated with such a Sudakov form factor. We will in the following see 
sevaral examples of this.

\vspace{2mm}

\emph{Formation time}

Assume that the charged particle moves along a curved and twisted trajectory. 
The emission probability is proportional to
\begin{equation}
\left| \int j(x) e^{i q x} \right|^2 .
\end{equation} 
From this expression we see that only short wavelengths can be sensitive to 
small deviations in the trajectory on a small timescale. For 
long wavelengths the integral gives the average over a larger region in 
space and time. This averaging implies that it takes a finite time to
determine that a photon has been emitted, which
can be formulated in terms of a (Landau-Pomeranchuk) formation time 
\begin{equation}
\tau \sim 1/q_\perp.
\label{formation-time}
\end{equation}

\emph{Ordered emissions} 

The result in eq.~(\ref{formation-time}) implies that in \eg an 
$e^{+}e^-$-annihilation event the photons with large $q_\perp$ are emitted
rapidly, directly after the annihilation, whereas those with smaller 
$q_\perp$ are emitted afterwards over an extended time period.
When many photons are emitted, the ordering in $q_\perp$ also corresponds 
to an ordering in time.

Looking for the first emission in time means looking for the emission
of the photon with largest $q_\perp$. The probability that no photon is emitted
with transverse momentum larger than $q_\perp$ is determined by a Sudakov 
form factor. Thus the probability for emitting a photon with transverse 
momentum $q_\perp$ as the first emission is given by 
\begin{equation}
dn \sim \alpha \frac{d x_1 d x_2}{(1-x_1)(1-x_2)} (x_1^2 + x_2^2)\cdot 
S(q_\perp) ,
\label{mu+mu-S}
\end{equation}
where the integral in $S(q_\perp)$ extends over all transverse momenta 
larger than $q_\perp$:

\begin{equation}
S(q_{\perp}) = \exp (-\int_{q'_{\perp} > q_{\perp}} d n) .
\label{sudakov-qt}
\end{equation} 

Similarly, if first a photon with transverse momentum $q_{\perp,1}$ is emitted,
a second emission of a softer photon with transverse momentum $q_{\perp,2}$ is 
associated with the form factor $S(q_{\perp,1},q_{\perp,2})$, defined by
\begin{equation}
S(q_{\perp,1},q_{\perp,2}) 
= \exp (-\int_{q_{\perp,2}}^{q_{\perp,1}} d q_{\perp} 
\frac{d n}{d q_{\perp}}).
\label{sudakov-1-2}
\end{equation} 
This process can be repeated to give a $q_\perp$-ordered cascade of emitted 
photons.

\vspace{2mm}

\emph{Spacelike cascades}

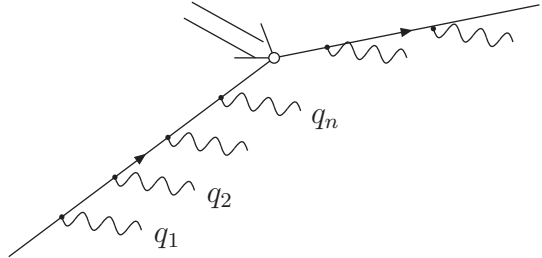
\begin{figure}
\begin{center}
\begin{picture}(200,96)(0,0)
\ArrowLine(0,0)(100,75) 
\ArrowLine(100,75)(200,95)
\Photon(20,15)(50,10){-3}{3} \Text(60,7)[]{ { \large{$q_1$} } }
\Photon(40,30)(70,25){-3}{3} \Text(80,22)[]{ { \large{$q_2$} } }
\Photon(60,45)(90,40){-3}{3}
\Photon(80,60)(110,55){-3}{3} \Text(120,52)[]{ { \large{$q_n$} } }
\Photon(120,79)(150,75){-3}{3}
\Photon(160,86)(190,81){-3}{3}
\Line(85,75)(100,75)
\Line(95,88)(100,75)
\Line(66,90)(93,75)
\Line(69,96)(96,81)
\BCirc(100,75){2}
\Vertex(20,15){1}
\Vertex(40,30){1}
\Vertex(60,45){1}
\Vertex(80,60){1}
\Vertex(120,79){1}
\Vertex(160,86){1}
\end{picture}
\end{center}
\caption{A spacelike photon cascade.}
\label{figure::dglapfoton}
\end{figure}

In spacelike cascades we have bremsstrahlung both as initial and as final 
state radiation, as indicated in fig.~\ref{figure::dglapfoton}. Final state 
radiation is very similar 
to the emission in $e^+e^-$-annihilation discussed above. The ordering due 
to formation time implies that also the initial radiation cascade is 
$q_\perp$-ordered:
\begin{equation}
q_{\perp, 1}^2 < q_{\perp, 2}^2 < \ldots\ <q_{\perp, n}^2 <Q^2
\label{spacelike-ordering}
\end{equation}

As mentioned earlier, neglecting recoils all photon emissions are 
uncorrelated, and therefore the ordering between them is not relevant. 
The ordering is only important when the recoils cannot be neglected,
and therefore the current is modified for the subsequent emissions.
However, as we will discuss more in the following, the ordering is
very essential in QCD, where the emitted radiation is not neutral, but
carries colour charge. This implies that an emitted gluon changes radically 
the current, which determines later (\ie softer) radiation. Therefore 
the order of the emissions becomes important, and is an 
essential feature of the evolution processes.

We also note that the separation between initial and final state radiation
in fig.~\ref{figure::dglapfoton} is gauge dependent. This separation is not 
determined by nature, but depends on the formalism used.

\section{Bremsstrahlung in QCD, \\ $e^+e^-$-annihilation}

\subsection{Gluon emission}

\begin{figure}
\begin{center}
\begin{picture}(200,73)(0,10) 
\Gluon(20,50)(100,50){4}{6} \Text(0,50)[]{ { \large{$p_1,\,\epsilon_1$} } }
\Gluon(100,50)(180,70){4}{6} \Text(200,70)[]{ { \large{$p_2,\,\epsilon_2$} } }
\Gluon(100,50)(150,20){4}{4} \Text(158,13)[]{ { \large{$q,\,\epsilon_3$} } }
\Vertex(100,50){1}
\end{picture}
\end{center}
\caption{Triple gluon vertex.}
\label{figure::treg}
\end{figure}
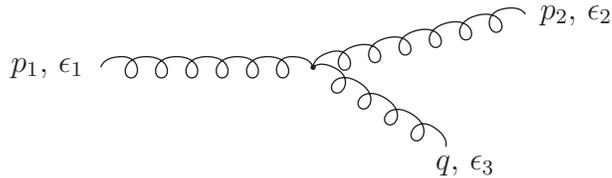

In QCD the gluons carry colour charge, and therefore radiate, or split into 
two gluons. The dominant interaction is the triple-gluon coupling (for 
notation \cf 
fig.~\ref{figure::treg}; $\epsilon_3$ denotes the polarization vector for 
the ``emitted'' gluon with momentum $q$):
\begin{equation}
g \left\{ [(p_1 + p_2)\epsilon_3] \,(\epsilon_1 \epsilon_2) -
[(q + p_1)\epsilon_2] \,(\epsilon_3 \epsilon_1) -
[(p_2-q)\epsilon_1] \,(\epsilon_2 \epsilon_3) \right\}
\label{tregluonkoppling}
\end{equation}
The first term is an eikonal contribution similar to the bremsstrahlung 
from a quark or the electromagnetic radiation from an electron in 
eq.~(\ref{soft}). The factor 
$\epsilon_1 \epsilon_2$ implies that the recoiling parent keeps its 
polarization, and it gives an emission density proportional to
\begin{equation}
\alpha_s \frac{d q_\perp^2}{q_\perp^2} d y .
\end{equation}
This term dominates for soft emissions. The second and third terms in 
eq.~(\ref{tregluonkoppling})
correspond to non-eikonal currents, in which the polarization vector for 
the emitted gluon, $\epsilon_3$, is multiplied by the polarization vector 
for the parent gluon, either before or after the recoil.

\subsection{Colour coherence}

\begin{figure}
\begin{center}
\begin{picture}(300,100)(0,0)

\Line(0,80)(35,50)
\Line(0,20)(35,50)
\Photon(35,50)(70,50){3}{5}
\Line(70,50)(80,57)
\Line(70,50)(120,10)
\Line(80,57)(120,90)
\Gluon(80,57)(122,73){2}{3}
\Text(78,61)[]{ { {$b$} } }
\Text(112,24)[]{ { {$\bar{b}$} } }
\Text(115,89)[]{ { {$r$} } }
\Text(130,77)[]{ { {$\bar{r}b$} } }
\Text(60,0)[]{ { {(a)} } }

\Line(150,80)(185,50)
\Line(150,20)(185,50)
\Photon(185,50)(220,50){3}{5}
\Line(220,50)(230,43)
\Line(220,50)(270,90)
\Line(230,43)(270,10)
\Gluon(230,43)(272,68){2}{3}
\Text(224,41)[]{ { {$\bar{r}$} } }
\Text(262,24)[]{ { {$\bar{b}$} } }
\Text(265,89)[]{ { {$r$} } }
\Text(281,70)[]{ { {$\bar{r}b$} } }
\Text(210,0)[]{ { {(b)} } }

\end{picture}
\end{center}
\caption{A quark and an antiquark emit gluons coherently by diagrams (a) and 
(b). In the angular region between the quark and the gluon the red and the 
antired charges radiate softer gluons approximately independently, forming a 
$r\bar{r}$ colour
dipole. At larger angles the emission from the red and the antired interfere 
destructively, and the quark-gluon system radiates as a single blue charge. 
Thus in this region the emission corresponds to a $b\bar{b}$ dipole.}
\label{figure::koherens}
\end{figure}
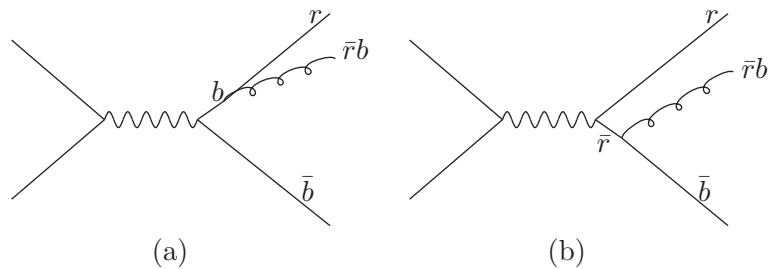

Study as an example the reaction in fig.~\ref{figure::koherens}a, 
where a blue-antiblue 
quark-antiquark pair is produced, and the blue quark emits a blue-antired 
gluon and becomes red. An identical final state can be obtained from
the diagram in fig.~\ref{figure::koherens}b, where the gluon is emitted from
an initially red antiquark, which becomes blue. The two diagrams interfere, 
and it is not possible to tell who is the parent to the emitted gluon.
It therefore makes sense to regard the quark and the antiquark as a colour 
dipole, which emits gluons coherently. The emission probability is given
by the corresponding expression for photon emission in eq.~(\ref{mu+mu-}), 
with $\alpha$ replaced by $\alpha_s$. 

In the region between the quark and the gluon, the 
separation of red and antired charges will give radiation of softer gluons. 
However, in directions further away from these particles the emission from 
red and antired interferes destructively. Thus in these directions the 
emission corresponds to the blue and antiblue charges of the initial 
$q\bar{q}$ pair, i.e. the emission from a $b\bar{b}$ colour dipole. This 
interference effect is generally called angular ordering \cite{angord}. In the 
restframe of the quark and gluon, the emission of softer gluons corresponds 
just to a $r\bar{r}$ dipole. Thus the emission of softer gluons corresponds 
to two \emph{independent} colour dipoles.

\subsection{Gluon cascades}

\begin{figure}
\begin{center}
\begin{picture}(200,91)(0,9) 
\Photon(30,50)(80,50){3}{5}
\Line(80,50)(160,90)
\Line(80,50)(160,10)
\Line(110,50)(165,50)
\Line(130,60)(164,65)
\Line(130,40)(164,32)
\Line(150,72)(167,78)
\end{picture}
\end{center}
\caption{A colour dipole cascade.}
\label{figure::dipolkaskad}
\end{figure}
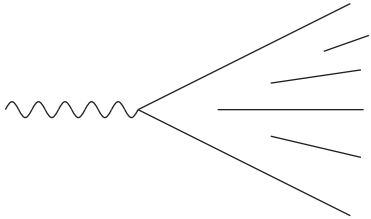

It is possible to describe a parton state in two equivalent (dual) ways 
\cite{dual}, either in terms of momenta, $q_i$, and spins, $s_i$, for the 
gluons, or 
in terms of momenta, $p_i$, and directions, $d_i$, for the dipoles. Thus 
gluon emission can be described as a process in which one dipole is split 
into two dipoles, which are split into four, etc., 
as shown in fig.~\ref{figure::dipolkaskad}. 
This formulation is used in the \emph{Dipole Cascade Model}
\cite{dual,dcm}. It 
is the basis for the \textsc{Ariadne} MC \cite{ariadne}, which has been very 
successfully applied to $e^+ e^-$ annihilation data, \cf \eg 
ref.~\cite{ariadnefit}.

The dipole formalism is also convenient for analytic studies of 
the properties of QCD cascades, and we also note that the resulting 
chain of colour dipoles
gives a very natural connection to the Lund String Fragmentation 
model~\cite{physrep}.

\section{DIS, DGLAP evolution}

\subsection{Notations}

The density of quarks and gluons within the proton is described by ``parton 
distribution functions'' $f_q(x)$ and $f_g(x)$, where $x$ is the fraction
of the proton momentum carried by the parton. We will use capital letters to
denote the densities in the variable $\ln 1/x$, and for simplicity we will
write $F(x)$ and $G(x)$ to denote $ x\!\cdot \!f_q(x)$ and 
$ x\!\cdot\! f_g(x)$ respectively.

The electron-proton cross section is expressed in terms of structure functions
$F_1$ and $F_2$. In the parton model these functions are related to each
other, and to the quark density. The exchanged photon is absorbed by a 
quark, which after the 
absorption stays on the mass shell. If $q_\gamma$ and $p_p$ denote 
the momenta of the photon and the proton, respectively, energy-momentum 
conservation is implemented by a delta-function $\delta(x - Q^2/2p_p \!\cdot 
\!q_\gamma)$,
where $Q^2=-q_\gamma^2$, and the $e\,p$ scattering cross section can 
(neglecting less important kinematic factors) be written
\begin{equation}
\frac{d^2 \sigma_{e\,p}}{d Q^2 \,d \!\ln 1/x} \sim \frac{\alpha^2}{Q^4} F_2(x) 
= \frac{\alpha^2}{Q^4} \sum_q e_q^2 \,x\,f_q(x) ,
\label{F2}
\end{equation}
where $x = x_{Bj} \equiv Q^2/2p_p \!\cdot \!q_\gamma$, the sum runs over quark 
species $q$,
and $e_q$ is the corresponding quark electric charge (in units of the 
elementary charge). The quantity $\frac{\alpha}{Q^2}F_2$
can also be interpreted as the total $\gamma^* p$ cross section.

In the distribution functions $F(x)$ and $G(x)$ the parton densities are 
integrated over 
transverse momenta. In the following we will also discuss ``non-integrated''
parton distributions, which depend also on $k_\perp$, and which we will denote
by curly letters, $\mathcal{F}(x,k_\perp^2)$ and $\mathcal{G}(x,k_\perp^2)$ 
for the quark and gluon densities respectively.

\subsection{Ordered ladders}

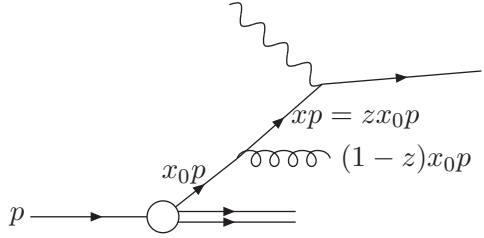
\begin{figure}
\begin{center}
\begin{picture}(200,92)(0,2)
\ArrowLine(10,10)(60,10) \Text(5,10)[]{ { {$p$} } }
\ArrowLine(60,10)(88,32)
\ArrowLine(88,32)(120,60)
\ArrowLine(120,60)(180,65)
\Photon(85,90)(120,60){3}{4}
\Gluon(88,32)(123,32){3}{4}
\ArrowLine(60,8)(110,8)
\ArrowLine(60,12)(110,12)
\BCirc(60,10){6}
\Text(68,26)[]{ { {$x_0 p$} } }
\Text(133,47)[]{ { {$x p=z x_0 p$} } }
\Text(120,32)[l]{ { {$(1-z) x_0 p$} } }
\end{picture}
\end{center}
\caption{A quark with momentum $p_0=x_0 p$ emits a gluon with 
momentum $(1-z)p_0$ before interacting with the photon.}
\label{figure::engluondis}
\end{figure}

In the parton model the structure function $F_2$ and the parton distributions 
are functions of $x$ only, but in QCD they also receive a dependence on $Q^2$.
Assume that a quark with momentum $p_0=x_0 p$ emits a gluon with 
momentum $(1-z)p_0$ before interacting with the photon, as shown in 
fig.~\ref{figure::engluondis}. This implies that
$x=Q^2/2pq_\gamma=z x_0$. The contribution to the cross section from the 
eikonal current is then determined by the emission probability
\begin{equation}
Prob. \sim \frac{4\alpha_s}{3\pi} \frac{d z}{1-z} 
\frac{d q_\perp^2}{q_\perp^2} \delta(x-z\,x_0) .
\label{z1pol}
\end{equation}

\begin{figure}
\begin{center}
\begin{picture}(300,100)(0,0)
\Line(25,0)(85,80)
\Line(85,80)(135,89)
\Photon(50,95)(85,80){3}{4}
\Gluon(40,20)(74,15){3}{4}
\Gluon(55,40)(87,35){3}{4}
\Gluon(70,60)(100,55){3}{4}
\Text(26,16)[]{ { {$p_0$} } }
\Text(41,36)[]{ { {$k_1$} } }
\Text(56,56)[]{ { {$k_2$} } }
\Text(71,16)[l]{ { {$q_1$} } }
\Text(84,36)[l]{ { {$q_2$} } }
\Line(160,0)(220,80)
\Line(220,80)(290,90)
\Photon(185,95)(220,80){3}{4}
\Gluon(175,20)(219,15){3}{5}

\Gluon(190,40)(222,35){3}{4}
\Gluon(205,60)(235,55){3}{4}
\Gluon(222,35)(249,35){3}{3}
\Gluon(222,35)(240,25){3}{2}
\Gluon(235,55)(255,63){3}{2}
\Gluon(235,55)(255,50){3}{2}
\Gluon(255,50)(288,55){3}{3}
\Gluon(255,50)(288,41){3}{3}
\Gluon(248,84)(265,75){-3}{2}
\Gluon(265,75)(285,75){2}{2}
\Gluon(265,75)(285,63){-2}{2}
\Gluon(269,87)(289,83){-2}{2}
\Gluon(197,50)(217,47){2}{3}
\Text(80,0)[]{ { {(a)} } }
\Text(230,0)[]{ { {(b)} } }
\end{picture}
\end{center}
\caption{(a) A chain formed by initial state radiation. (b) To describe 
properties of exclusive final states final state radiation has to be added.
Note that a soft final state gluon also can be emitted from a virtual link,
if the recoil is small. Such emissions, like the one emitted from the link 
$k_2$ in this figure, will be further discussed in section 5.5.}
\label{figure::ISR}
\end{figure}
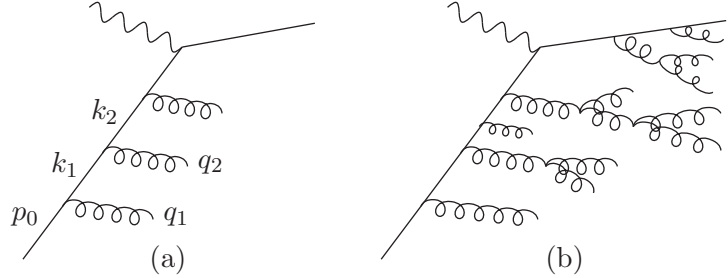

When many gluons are emitted as in fig.~\ref{figure::ISR}a we again obtain 
an ordered emission determined by the formation times:
\begin{equation}
q_{\perp 1}^2 < q_{\perp 2}^2 < \ldots\ q_{\perp n}^2 < Q^2
\label{ordningdglap}
\end{equation}
Thus the contribution to the structure function $F$ from a chain with
$n$ links is given by a product of $n$ factors
\begin{equation}
\frac{4\alpha_s}{3\pi} \frac{d z_i}{1-z_i} 
\frac{d q_{\perp i}^2}{q_{\perp,i}^2} ,
\label{dglaplink}
\end{equation}
where $x=(\prod^n z_i)x_0$, and the transverse momenta satisfy the ordering 
condition in eq.~(\ref{ordningdglap}).

Although many gluons are emitted in the process in fig.~\ref{figure::ISR}a,
there is still only
\emph{one} quark, which can be hit by the photon. When $Q^2$ is large, the 
many gluons, which can be emitted before the $\gamma q$ interaction, imply a 
reduced probability to collide \emph{without} gluon emission. The factor in 
eq.~(\ref{dglaplink}), which determines the probability to emit a gluon, must 
then be multiplied by a \emph{Sudakov form factor}, $S$, which describes the 
probability that no emission has occurred between $q_{\perp,i-1}$ and 
$q_{\perp,i}$. Thus we have to make the following replacement in 
eq.~(\ref{dglaplink}):
\begin{equation}
\frac{1}{1-z_i} \frac{d q_{\perp, i}^2}{q_{\perp, i}^2} \rightarrow
\frac{\theta(1-\epsilon -z_i)}{1-z_i} \frac{d q_{\perp,i}^2}{q_{\perp,i}^2}
S(q_{\perp,i}^2, q_{\perp,i-1}^2)
\label{sudakovvertex} 
\end{equation}
where
\begin{equation}
S(q_{\perp,i}^2, q_{\perp,i-1}^2) = \exp\left[ 
- \int_{q_{\perp,i-1}^2}^{q_{\perp,i}^2} \frac{4\alpha_s}{3\pi}
\frac{d q_{\perp}^2}{q_{\perp}^2}
\int_0^{1-\epsilon} \frac{d z}{1-z}\right]
\label{sudakovii-1} 
\end{equation}
We have here introduced a cutoff $\epsilon$, which 
in a MC must be kept different from zero, since 
$\epsilon \rightarrow 0$ implies that $S \rightarrow 0$. A small 
$\epsilon$ gives a more accurate result, but also a slower program.

\subsection{DGLAP evolution for large $x$}

Summarizing the results in the previous subsection, and summing over the 
number of links, $n$, in the chain, we find the following expression for the 
structure function:
\begin{eqnarray}
F \sim \sum_n \prod_i^n \left\{ \int \frac{4\alpha_s}{3\pi} \frac{d z_i}{1-z_i}
\frac{d q_{\perp i}^2}{q_{\perp,i}^2} S(q_{\perp,i}^2, q_{\perp,i-1}^2)
\theta(q_{\perp,i} - q_{\perp,i-1})
\right\} \times \nonumber \\
\times \delta(x-\prod_j^n z_j \cdot x_0) \theta(Q^2 - q_{\perp,n}^2)
\label{dglapchain}
\end{eqnarray}
We note that in this result most of the $z$-values are close to 1. We
also note that, due to transverse momentum conservation
 and the ordering in eq.~(\ref{ordningdglap}), the transverse momentum of
the (quasi)real emissions, $q_{\perp,i}$, are approximately equal to the
transverse momentum of the virtual links, which in fig.~\ref{figure::ISR}a are
denoted $k_{\perp,i}$. Thus the variables $q_{\perp,i}$ in 
eq.~(\ref{dglapchain})
can to the given accuracy be replaced by the link variables $k_{\perp,i}$.

It is also possible to include the effects of the spin of the quarks by the 
exchange
\begin{equation}
\frac{1}{1-z}\, \rightarrow \,\frac{\frac{1}{2} (1+z^2)}{1-z}
\label{kvarkspin}
\end{equation}
Here the numerator  corresponds to (half) the factor $(x_1^2 + x_2^2)$ in 
eq.~(\ref{mu+mu-}), and 
does not change the behaviour near the dominating singularities at $z_i=1$.

Taking the derivative of eq.~(\ref{dglapchain}) with 
respect to $\ln Q^2$ gives the following differential-integral equation 
(the DGLAP evolution equation):
\begin{equation}
\frac{\partial F(x,Q^2)}{\partial \ln Q^2} = \frac{4\alpha_s}{3\pi} 
\int dz \,dx' \hat{P}(z) F(x',Q^2) \delta(x-zx') .
\label{DGLAP}
\end{equation}
It has here been possible to take the limit $\epsilon \rightarrow 0$, and 
thus remove the dependence on the cutoff. The 
``splitting function'' $P$ in eq. (\ref{kvarkspin}) is then replaced by 
$\hat{P}$ :
\begin{equation}
P= \frac{\frac{1}{2} (1+z^2)}{1-z} \,\rightarrow \,
\hat{P}=\frac{\frac{1}{2} (1+z^2)}{(1-z)_+}+\frac{3}{2} \delta(1-z) ,
\label{phat}
\end{equation}
where the definition of the ``+ prescription'' is given by
\begin{equation}
\int \frac{d z}{(1-z)_+} \cdot f(z) \equiv 
\int d z \,\frac{f(z) - f(1)}{(1-z)}
\label{plus-prescription}
\end{equation}
for an arbitrary function $f(z)$. We note in particular that this definition 
implies that 
\begin{equation}
\int \hat{P}(z) dz = 0 ,
\label{phatnorm}
\end{equation} 
which guarantees that the number of quarks is conserved.

\subsection{Exclusive final states}

We want to emphasise that eq.~(\ref{DGLAP}) only describes the probability for
an interaction, that is the total photon-proton cross section expressed in 
terms of the structure function $F_2$. To find the properties of exclusive 
final states it is necessary to add 
final state radiation within angular ordered regions, as indicated in 
fig.~\ref{figure::ISR}b.
Just as the final state emission in $e^+e^-$-annihilation, these emissions
do not affect the total cross section, and should therefore also be associated 
with appropriate Sudakov form factors.

\subsection{Gluon rungs and small $x$}

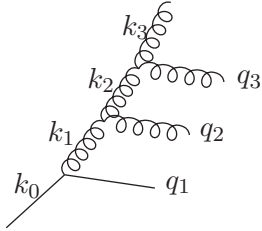
\begin{figure}
\begin{center}
\begin{picture}(120,90)(0,0)
\Line(18,0)(40,20)
\Gluon(40,20)(55,40){3}{4}
\Gluon(55,40)(68,60){3}{4}
\Gluon(68,60)(80,85){3}{4}
\Line(40,20)(74,15)
\Gluon(55,40)(87,35){3}{4}
\Gluon(68,60)(100,55){3}{4}
\Text(26,16)[]{ { {$k_0$} } }
\Text(39,36)[]{ { {$k_1$} } }
\Text(54,56)[]{ { {$k_2$} } }
\Text(67,76)[]{ { {$k_3$} } }
\Text(71,16)[l]{ { {$q_1$} } }
\Text(84,36)[l]{ { {$q_2$} } }
\Text(98,56)[l]{ { {$q_3$} } }
\end{picture}
\end{center}
\caption{A gluon ladder starting from a quark $k_0$. The quasireal emissions 
have momenta $q_i$, while the virtual links are denoted $k_i$.}
\label{figure::gluonkedja}
\end{figure}

Also gluons can emit gluon radiation. Study the emission of gluon $q_2$ in
fig.~\ref{figure::gluonkedja}.
With the notation in this figure the \emph{eikonal current} corresponding
to the first term in eq.~(\ref{tregluonkoppling}) gives
$g \,(k_1+k_2)\epsilon_{q_2}\,(\epsilon_{k_1}\epsilon_{k_2})$.
This term gives a contribution similar to that for emission from a quark
in eqs.~(\ref{dglaplink}, \ref{sudakovvertex}). Including now
only the term singular at $z=1$ we get
\begin{equation}
\frac{3\alpha_s}{\pi} \frac{d z_i}{1-z_i} 
\frac{d q_{\perp i}^2}{q_{\perp,i}^2} \theta(q_{\perp,i} - q_{\perp,i-1})
\cdot S.
\label{eikonal}
\end{equation}

The spin factor $\epsilon_{k_1}\epsilon_{k_2}$ implies that the gluon spin 
is conserved along the vertical line in the ladder in 
fig.~\ref{figure::gluonkedja}. We also note 
that, just as for quark emission, the pole at $z=1$ implies that $z$-values 
close to 1 dominate. Thus gluon $k_2$ can be regarded as identical to the 
recoiling parent $k_1$; it inherits both the spin and most of the energy, 
while the gluon $q_2$ has to be regarded as a newborn soft gluon.

The \emph{non-eikonal} contribution to the current is proportional to
the remaining terms in eq.~(\ref{tregluonkoppling}),
$-(k_1+q_2)\epsilon_{k_2}\,(\epsilon_{k_1}\epsilon_{q_2})
- (q_2-k_2)\epsilon_{k_1}\,(\epsilon_{k_2}\epsilon_{q_2})$.
Here the first term dominates (for a conventional gauge choice), and gives 
a contribution $\sim dz_i/z_i$:
\begin{equation}
\frac{3\alpha_s}{\pi} \frac{d z_i}{z_i} 
\frac{d q_{\perp i}^2}{q_{\perp,i}^2} \theta(q_{\perp,i} - q_{\perp,i-1}).
\label{non-eikonal}
\end{equation}

Thus the ``soft'' gluon goes up the ladder, and it is the (quasireal) 
gluon $q_2$, which takes over both the spin and most of the energy of the 
parent $k_1$. 
The soft daughter does not replace the 
mother, and therefore there is \emph{no Sudakov form factor} in leading order.

We note that soft gluon links can also be emitted from \emph{quark} legs, 
as also indicated in fig.~\ref{figure::gluonkedja}. The probability for this 
process has a different colour factor, and is proportional to
\begin{equation}
\frac{4\alpha_s}{3\pi} \frac{d z}{z} 
\frac{d q_{\perp}^2}{q_{\perp}^2} ,
\label{quark-vertex}
\end{equation}
and also in this contribution there is no Sudakov form factor to leading order.

For gluonic chains the sum of the expressions in eqs.~(\ref{eikonal}) and 
(\ref{non-eikonal}) replace the 
corresponding factor in eq.~(\ref{dglapchain}). The non-eikonal terms in 
eq.~(\ref{non-eikonal}) give the dominant 
contributions for small $z$-values, and therefore also for small 
$x=(\prod z_i)\,x_0$. Thus gluon ladders are most important for the growth 
of the structure functions for small $x$, and to leading order in
$\ln 1/x$ and $\ln Q^2$ the gluon density can be written
\begin{equation}
G(x,Q^2) \sim \sum_n \prod_i^n \left\{ \int \frac{4\alpha_s}{3\pi} \frac{d z_i}{z_i}
\frac{d q_{\perp i}^2}{q_{\perp,i}^2}
\theta(q_{\perp,i} - q_{\perp,i-1})
\right\} 
\, \delta(x- \prod_j^n z_j \cdot x_0) \theta(Q^2 - q_{\perp,n}^2)
\label{smallxchain}
\end{equation}
The gluon ladder may start from an initial quark with the coupling in 
eq.~(\ref{quark-vertex}), and in DIS it must also have a quark link at the end,
as the photon only couples to the electrically charged quarks. It is, however, 
the properties of the dominating gluonic ladder, which determines 
the asymptotic growth rate for small $x$.

\subsection{Double Leading Log approximation}

Assume that both $Q^2$ and $1/x$ are very large. We introduce the notation
\begin{eqnarray}
x_i &=&\frac{k_i}{p_{\mathrm{proton}}} = \prod_j^i z_j \\
\bar{\alpha} &\equiv& \frac{3\alpha_s}{\pi}  \\
\kappa_i &\equiv& \ln (q_{\perp,i}^2/\Lambda^2)  \\
l_i &\equiv& \ln (1/x_i)
\label{defkappal}
\end{eqnarray}
For a fixed coupling $\bar{\alpha}$, we then find from eq.~(\ref{smallxchain})
\begin{eqnarray}
G &\sim \sum_n \left\{ \prod_i^n \int^{\ln Q^2} \bar{\alpha} d\kappa_i 
\theta(\kappa_i - \kappa_{i-1})   
\cdot \prod_i^n \int^{\ln 1/x} dl_i 
\theta(l_i - l_{i-1}) \right\}  =        \nonumber \\
&=\sum_n \bar{\alpha}^n \cdot \frac{(\ln Q^2)^n}{n!}\cdot\frac{(\ln 1/x)^n}{n!}
= \nonumber \\
&= I_0(2\sqrt{\bar{\alpha} \ln Q^2 \ln 1/x})
\sim \exp\left(2\sqrt{\bar{\alpha} \ln Q^2 \ln 1/x}\right) ,
\label{DLL}
\end{eqnarray}
where $I_0$ is a modified Bessel function.
This result corresponds to the Double Leading Log (DLL) approximation.
 
For a running $\alpha_s$ we define the parameter $\alpha_0$ by the relation
\begin{equation}
\bar{\alpha} \equiv \frac{\alpha_0}{\ln (q_\perp^2/\Lambda^2)}.
\label{alfabar}
\end{equation}
We then have instead of $d q_{\perp,i}^2 /q_{\perp,i}^2 = d \kappa_i$ a factor 
$d\kappa_i/\kappa_i = d \ln \kappa_i$, which gives the result
\begin{equation}
G \sim  \exp\left(2\sqrt{\alpha_0 \cdot \ln \ln Q^2 \cdot \ln 1/x}\right) .
\label{DLLlopande}
\end{equation}

\section{Small $x$, the BFKL region}

\subsection{Non-ordered ladders}

Now assume that $Q^2$ is not large, while $x$ is still kept small. In this 
case the $q_\perp$-ordered phase space is small. Therefore 
$q_\perp$-non-ordered contributions are important, even if they are 
suppressed. 

We mentioned after eq.~(\ref{dglapchain}) that for a $q_\perp$-ordered chain 
the  transverse momentum of
the (quasi)real emissions, $q_{\perp,i}$, are approximately equal to the
transverse momentum of the virtual links, denoted $k_{\perp,i}$ 
in figs.~\ref{figure::ISR} and \ref{figure::gluonkedja}. 
Therefore the variables $q_{\perp,i}$ in 
eqs.~(\ref{dglapchain}) and (\ref{smallxchain}) could be replaced by the 
link variables $k_{\perp,i}$. This is no longer the case
for non-ordered chains. Transverse momentum conservation implies that
$\mathbf{q}_{\perp,i} = \mathbf{k}_{\perp,i-1} -  \mathbf{k}_{\perp,i}$, and 
therefore we have, provided $\mathbf{k}_{\perp,i-1}$ and  
$\mathbf{k}_{\perp,i}$
are not approximately equal,
\begin{equation}
q_{\perp,i}^2 \approx \max(k_{\perp,i}^2, k_{\perp,i-1}^2).
\label{q-k-samband}
\end{equation}
Emissions for which $\mathbf{k}_{\perp,i-1} \approx \mathbf{k}_{\perp,i}$
will be discussed in subsection 5.5. They give no contribution to the
structure functions, and can be treated as final state radiation. For
other emissions eq.~(\ref{q-k-samband})
implies that if \eg $k_{\perp,i}$ is larger than the neighbouring links
($k_{\perp,i-1}$ and $k_{\perp,i+1}$), then $q_{\perp,i}\approx q_{\perp,i+1}
\approx k_{\perp,i}$. If on the other hand $k_{\perp,i}$ is smaller than its
neighbours, then its value is not close to any of the transverse momenta 
$q_{\perp}$. Thus the quasi-real momenta $q_{\perp,i}$ are approximately 
determined by the virtual links $k_{\perp,i}$, but the reverse is not true
(without the knowledge of the azimuthal angles of all the vectors 
$\mathbf{q}_{\perp,i}$).
To specify the chain we therefore have to specify the link variables 
$k_{\perp,i}$,
and the distinction between $q_{\perp,i}$ and $k_{\perp,i}$ is essential.

Classical bremsstrahlung due to a short impulse during a time 
$\Delta t \sim 1/Q$ contains the factor

\begin{equation}
\sim \left| \int_0^{1/Q} dt \e^{i\omega t} \right|^2 = \left| \frac{1}{\omega}
(e^{i\omega /Q} -1) \right|^2 
=\left\{ \begin{array}{ll}
\frac{1}{Q^2},  &\mathrm{for}\,\,\, \omega << Q \\
\sim \frac{1}{\omega^2},  
&\mathrm{for}\,\,\, \omega >> Q
\end{array} \right.
\label{impuls}
\end{equation}
Thus for $\omega >> Q$ there is a relative suppression by a factor
$Q^2 / \omega^2$. In a relativistic calculation we should make the replacement 
$\omega \rightarrow q_\perp$. \emph{We then find that for $q_\perp^2 > Q^2$ 
the emission is not totally excluded; it is only reduced by a suppression
factor $Q^2 / q_\perp^2$.}

Assume that we have an ordered chain up to the last link with $k_{\perp,n}^2
\approx q_{\perp,n}^2$, which can be smaller or larger than $Q^2$. 
We then find for these two cases:
\begin{itemize}
\item
$k_{\perp,n}^2 < Q^2$.

In this ordered case we have 
$Q^2 > q_{\perp,n}^2 \approx k_{\perp,n}^2 >k_{\perp,n-1}^2 $, and 
no extra suppression. According to 
eqs.~(\ref{dglapchain}) and (\ref{smallxchain}) the contribution to the 
structure function from these chains contain a factor $1/q_{\perp,n}^2$.
Therefore we have
\begin{equation}
\sigma \sim \frac{1}{Q^2}\cdot F \sim  
\frac{1}{Q^2 q_{\perp,n}^2}.
\label{slutupp}
\end{equation}

\item
$k_{\perp,n}^2 > Q^2$.

This implies
$Q^2 < q_{\perp,n}^2 \approx k_{\perp,n}^2 >k_{\perp,n-1}^2 $,
and in this situation there is a 
suppression factor $Q^2/ q_{\perp,n}^2$. Thus we have
\begin{equation}
\sigma \sim \frac{1}{Q^2}\cdot \frac{1}{q_{\perp,n}^2} 
\cdot \frac{Q^2}{q_{\perp,n}^2}\sim  
\frac{1}{k_{\perp,n}^4}.
\label{slutner}
\end{equation}

\end{itemize}

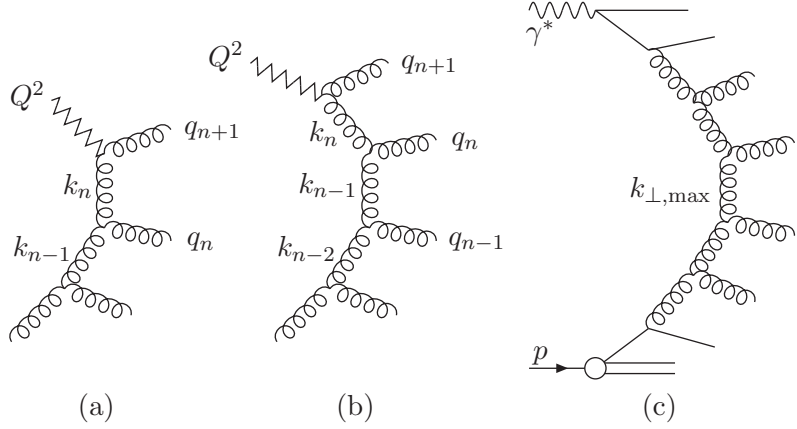
\begin{figure}
\begin{center}
\begin{picture}(300,165)(0,0)
\Gluon(0,25)(20,45){3}{4}
\Gluon(20,45)(35,68){3}{4} \Text(30,60)[r]{ { {$k_{n-1}$} } }
\Gluon(35,68)(35,96){3}{4} \Text(38,84)[r]{ { {$k_{n}$} } }
\Gluon(20,45)(45,35){3}{4}
\Gluon(35,68)(60,63){3}{4} \Text(59,63)[l]{ { {$q_{n}$} } }
\Gluon(35,96)(60,105){3}{4} \Text(58,105)[l]{ { {$q_{n+1}$} } }
\ZigZag(15,116)(35,96){3}{5} \Text(20,118)[r]{ { {$Q^2$} } }
\Text(32,0)[]{ { {(a)} } }

\Gluon(100,25)(120,45){3}{4}
\Gluon(120,45)(135,68){3}{4} \Text(130,60)[r]{ { {$k_{n-2}$} } }
\Gluon(135,68)(135,96){3}{4} \Text(138,84)[r]{ { {$k_{n-1}$} } }
\Gluon(117,118)(135,96){3}{4} \Text(131,104)[r]{ { {$k_{n}$} } }
\Gluon(120,45)(145,35){3}{4}
\Gluon(135,68)(160,63){3}{4} \Text(158,63)[l]{ { {$q_{n-1}$} } }
\Gluon(135,96)(160,101){3}{4} \Text(159,101)[l]{ { {$q_{n}$} } }
\Gluon(117,118)(142,130){3}{4} \Text(140,130)[l]{ { {$q_{n+1}$} } }
\ZigZag(90,130)(117,118){3}{5} \Text(95,133)[r]{ { {$Q^2$} } }
\Text(130,0)[]{ { {(b)} } }

\ArrowLine(195,15)(220,15) \Text(200,20)[]{ { {$p$} } }
\Line(220,13)(250,13)
\Line(220,17)(250,17)
\Line(220,15)(240,30) \Line(240,30)(265,23)
\Gluon(240,30)(257,50){3}{4} \Gluon(257,50)(280,40){3}{4}
\Gluon(257,50)(270,70){3}{4} \Gluon(270,70)(295,65){3}{4}
\Gluon(270,70)(270,95){3}{4} \Text(272,82)[r]{ { {$k_{\perp,\mathrm{max}}$} } }
\Gluon(257,115)(270,95){3}{4} \Gluon(270,95)(295,100){3}{4}
\Gluon(240,135)(257,115){3}{4} \Gluon(257,115)(280,125){3}{4}
\Line(220,150)(240,135) \Line(240,135)(265,140)
\Line(220,150)(255,150)
\Photon(195,150)(220,150){3}{4} \Text(200,142)[]{ { {$\gamma^*$} } }
\BCirc(220,15){4}
\Text(245,0)[]{ { {(c)} } }
\end{picture}
\end{center}
\caption{(a) A situation where $k_{\perp,n}^2$ is larger than the virtuality 
$Q^2$ of the probe can be interpreted as a boson-gluon-fusion event, 
\emph{i.e.} 
a hard subcollision between the probe and gluon $k_{n-1}$. (b) If 
$k_{\perp,n-1}^2$ is the largest virtuality, we have a hard subcollision 
between gluons $k_{n-2}$ and $k_{n}$. (c) When the gluon with the
largest transverse momentum is located in the middle of the chain, it 
corresponds to a hard subcollision between a proton and a resolved photon.}
\label{figure::hard}
\end{figure}

In the latter case we see that the process can be interpreted as a hard 
subcollision between the probe and the
gluon $k_{n-1}$, \cf fig.~\ref{figure::hard}a. (Instead of a photon we imagine 
here a hypothetical colour neutral probe, which can interact directly with 
gluons.) We recognize the expected result from a hard 
scattering with momentum exchange $t=-k_{\perp,n}^2$, which is then 
the largest virtuality in the process.

Assume now that we have a situation where the last but one gluon link,
$k_{n-1}$, has the largest transverse momentum, which implies: 

\begin{itemize}
\item
$Q^2 < k_{\perp,n}^2 < q_{\perp,n-1}^2 \approx k_{\perp,n-1}^2
\approx q_{\perp,n-2}^2 > k_{\perp,n-2}^2$.
\end{itemize}
We expect here a suppression factor 
$Q^2/q_{\perp,n-1}^2\approx Q^2/k_{\perp,n-1}^2$. 
This factor can be written as a
product of two separate factors, one for each step downwards in $k_\perp$:
\begin{equation}
\frac{Q^2}{k_{\perp,n-1}^2} =\frac{k_{\perp,n}^2}{k_{\perp,n-1}^2} \cdot 
\frac{Q^2}{k_{\perp,n}^2} 
\label{dubbelslutstegner}
\end{equation}
As illustrated in fig.~\ref{figure::hard}b, this process can be interpreted 
as a hard subcollision between the links $k_n$ (now coming from above) 
and $k_{n-2}$, with 
momentum transfer $k_{\perp,n-1}^2$. The outgoing gluons have approximately
equal but opposite transverse momenta, $q_{\perp,n-1}$ and $q_{\perp,n}$,
and the cross section satisfies
the expected relation $\sigma \sim 1/k_{\perp,n-1}^4$.

These results can be generalized to the situation in fig.~\ref{figure::hard}c,
where $k_\perp$ increases continously from the proton, up to a 
maximum value $k_{\perp,\mathrm{max}}$, and then decreases in $k_\perp$ down to the 
virtuality, $Q^2$, of the photon. This process has a weight with a factor 
$1/k_{\perp,i}^2$
for every links except $k_\mathrm{max}$, which instead gives a factor 
$1/k_{\perp,\mathrm{max}}^4$.
We note here that, although upwards and downwards steps are treated 
differently, the final result is \emph{symmetric} in the sense that we could 
equally well
have started the chain from the photon end, and proceeded towards the proton.
Therefore it is identical to the result obtained from a 
DGLAP evolution for both the proton in one end and a \emph{resolved photon} 
in the other end, up to a 
central hard scattering, with momentum transfer $k_{\perp,\mathrm{max}}$.

With enough energy it is apparently also possible to have ladders in which 
the transverse 
momentum goes up and down with two or more local maxima. When expressed in the
virtual links $k_\perp$, each step downwards
corresponds to a suppression factor $k_{\perp, i}^2 / k_{\perp, i-1}^2$. 
The net result is a factor $1/k_{\perp}^{4}$ for every local maximum 
$k_\perp$, but no $k_\perp$-power for links corresponding to a local minimum.

We formulated the result in terms of the link variables $k_{\perp,i}^2$,
because these could be interpreted as independent variables.
Transverse momentum conservation implies that the quantities  $q_{\perp,i}$
are not independent, but constrained by the relation
$q_{\perp,\mathrm{max}}^2 \approx q_{\perp,\mathrm{max}+1}^2$, and they do not fix 
the value of $k_{\perp}^2$ for a link which represents a local minimum. 
We note, however, that the weight for the chain corresponds
exactly to the product $\prod_i^n q_{\perp,i}^{-2}$. The factor
$q_{\perp,\mathrm{max}}^{-2} \cdot q_{\perp,\mathrm{max}+1}^{-2}$ gives the factor
$1/k_{\perp,\mathrm{max}}^{4}$, and the $k_{\perp}$ for 
a local minimum, which does not equal any $q_\perp$, does not appear in the 
weight.

We further note that, if the azimuthal angles are not averaged over, but 
properly accounted for,
we have $d^2 k_{\perp,i} = d^2 q_{\perp,i}$. As a consequence we see that 
our result exactly corresponds to the ordered result in 
eq.~(\ref{smallxchain}),
if we make the replacement 
\begin{equation}
\frac{d q_{\perp,i}^2}{q_{\perp,i}^2} \rightarrow 
\frac{d^2 q_{\perp,i}}{\pi q_{\perp,i}^2} ,
\label{d2q}
\end{equation}
omit the ordering $\theta$-functions and instead include a 
factor \\ 
$\theta(q_{\perp,i} - \min(k_{\perp,i}, k_{\perp,i-1}))$. This factor
excludes soft emissions for which $\mathbf{k}_{\perp,i} \approx 
\mathbf{k}_{\perp,i-1}$, as mentioned above and is further discussed
in section 5.5.
Thus the expression
\begin{equation}
G \sim \sum_n \prod_i^n \left\{ \int \frac{4\alpha_s}{3\pi} \frac{d z_i}{z_i}
\frac{d^2 q_{\perp i}}{\pi q_{\perp,i}^2}
\theta(q_{\perp,i} - \min(k_{\perp,i}, k_{\perp,i-1}))
\right\}  \delta(x- \prod_j^n z_j \cdot x_0)
\label{bfklchain}
\end{equation}
gives a proper description, both for large $Q^2$ and ordered chains, and for
limited $Q^2$ when non-ordered chains are important. To get the non-integrated 
distribution function $\mathcal{G}$, we just have to add a factor 
$\delta(\mathbf{k}_{\perp,n} + \sum_j^n \mathbf{q}_{\perp,j})$, which
follows from conservation of transverse momentum.

This result also demonstrates the symmetry discussed above.
An important consequence of this symmetry is that the formalism also
is suitable for describing hard subcollisions in hadronic collisions.
This will be further discussed in section 6.

\subsection{Effective phase space}

With more than one local maximum in the $k_\perp$ chain, we find for every 
step down from $k_{\perp, i-1}^2$ to $k_{\perp, i}^2$ a suppression factor 
$k_{\perp, i}^2 / k_{\perp, i-1}^2$. Expressed in the logarithmic variable 
$\kappa = \ln k_\perp^2$, this corresponds to a factor 
$\exp[-(\kappa_{i-1} - \kappa_i)]$. This 
implies that \emph{the effective range allowed for downward steps corresponds 
to approximately one unit in $\kappa$}. Consequently we find instead of the 
phase space limits in eq.~(\ref{ordningdglap}), the following boundaries
(replacing $q_{\perp, i}^2$ by $k_{\perp, i}^2$, which are the relevant
variables  for non-ordered chains)
\begin{equation}
\ln k_{\perp, i}^2 \gtaet \ln k_{\perp, i-1}^2 - 1.
\label{bfklfasrum}
\end{equation}
This modification changes the DLL result in eqs.~(\ref{DLL}) and 
(\ref{DLLlopande}) in a 
qualitative way, as described in the following subsections. 

\subsection{Fixed $\alpha_s$}

For a fixed $\alpha_s$, and writing $\kappa$ for $\ln Q^2$,
 we then find for the transverse momentum integrals in
 eq.~(\ref{DLL}) instead of

\begin{equation}
\int_0^\kappa \prod_i^n d\kappa_i \theta(\kappa_i - \kappa_{i-1}) = 
\frac{\kappa^n}{n!}, \,\,\,
\,\,\, (\kappa=\ln Q^2)
\label{fixalfa}
\end{equation}
the following result 
\begin{equation}
\int_0^\kappa \prod_i^n d\kappa_i \theta(\kappa_i - \kappa_{i-1} - 1) \approx 
\frac{(\kappa+n)^n}{n!} 
\label{fixalfasmallx}
\end{equation}

When $\kappa$ is very large we recover the DLL result in eq.~(\ref{DLL}), 
but for smaller values of $\kappa$ we find instead using Sterling's formula
\begin{equation}
\kappa \,\,\mathrm{small}\,\,\Rightarrow \,\,\frac{(\kappa+n)^n}{n!} 
\sim \frac{n^n}{n!} \sim \e^n
\label{kappalitet}
\end{equation}
which implies
\begin{equation}
G \sim \sum_n \frac{[ \bar{\alpha}\, \e \ln(1/x)]^n}{n!} = 
\e^{\e \,\bar{\alpha} \ln(1/x)} = \frac{1}{x^\lambda}
\label{xlambda}
\end{equation} \\
with 
\begin{equation}
\lambda = \e \,\bar{\alpha} \approx 2.72 \,\bar{\alpha}.
\label{approxlambda}
\end{equation}

To obtain the non-integrated distribution $\mathcal{G}(x, k_\perp^2)$,
we add the factor $\delta(\mathbf{k}_\perp-\mathbf{k}_{\perp,n})$,
but this does not modify the qualitative result, if $\kappa$ now denotes $\ln
k_\perp^2$. 

The result in eq.~(\ref{xlambda}) is relevant for
$\ln k_\perp^2 < \e \,\bar{\alpha} \ln(1/x)$. In this range
the chain corresponds to a random walk in $\ln k_\perp^2$.
Our result should be compared with the result from the leading order 
BFKL equation, which gives
\begin{equation}
\lambda = 4\ln2 \,\bar{\alpha} \approx 2.77 \,\bar{\alpha}.
\label{bfkllambda}
\end{equation}
We see that this simple semiclassical picture describes the essential 
features of BFKL evolution.

\subsection{Running $\alpha_s$}

For a running $\alpha_s$ the steps in transverse momentum are determined 
by the factors
\begin{equation}
\alpha_0 \frac{d \ln k_{\perp,i}^2}{\ln k_{\perp,i}^2} 
= \alpha_0 d(\ln \ln k_{\perp,i}^2)
\label{lopsteg}
\end{equation}
When $k_\perp$ is large, one extra unit in $\ln k_{\perp}^2$ corresponds to 
a very small extra space in $\ln \ln k_{\perp}^2$. This implies that, once
the chain has reached large $k_\perp$-values, it is very difficult to come 
down again. Therefore dominant chains will contain
an initial part with low  $k_\perp$ and steps up and down, and a 
second (DGLAP-like) part with increasing $k_\perp$ up to the final $k_\perp$- 
and $x$-values. As a consequence the structure functions can be well described 
by DGLAP evolution from adjusted input distributions $f_0(x,k_{\perp 0}^2)$, 
which grow for small values of $x$.

For further discussions of the results in this and the previous sections,
see ref.~\cite{simplemodel}.

\subsection{Exclusive final states}

As discussed above in connection with large-$Q^2$ phenomena, final state 
radiation must be added, in order to describe the properties of exclusive 
final states. The kinematic regions in which such emission should be allowed 
is in general dependent on the formalism used. The 
quasireal gluons, denoted $q_i$ in figs.~\ref{figure::ISR}
and \ref{figure::gluonkedja}, are allowed to emit final
state radiation in angular ordered regions, but besides within these regions 
further soft emissions may also be possible.

In all discussions of the results in the previous sections we 
studied chains in which at every vertex
the emitted $q_\perp$ approximately equals the larger of the 
adjoining $k_\perp$. We could also imagine the emission of a softer gluon,
with $q_{\perp,i} << k_{\perp,i-1}$, which then implies 
$\mathbf{k}_{\perp,i}\approx \mathbf{k}_{\perp,i-1}$. From the discussion about formation time 
in section 2.2, such emissions should take place at a longer time scale,
after formation of the interaction chain in figure \ref{figure::ISR}a or
\ref{figure::gluonkedja}.
They should therefore 
not influence the reaction probability, \ie the structure function, but only
affect the properties of exclusive final states. Consequently
they may be treated as final state radiation, and must be associated with 
appropriate Sudakov form factors. This is indeed the result obtained in
a perturbative calculation, in which such emissions are compensated by virtual 
corrections, and in the BFKL formalism this effect is described by treating 
the link gluons as \emph{Reggeized gluons}. Such final state radiation
may be emitted both from gluon and quark links, and one example
was indicated in fig.~\ref{figure::ISR}b.

These emissions are also treated as final state radiation in 
the \emph{Linked Dipole Chain} (LDC) model for DIS \cite{LDC}, 
which will be further discussed in the next section.
We note, however, that the separation between initial and final state
radiation is not defined by nature, but depends on the formalism used.
This separation is defined in a different way in the CCFM formalism 
\cite{ccfm}, in which some emissions, for which 
$\mathbf{k}_{\perp,i} \approx \mathbf{k}_{\perp,i-1}$, are treated as initial 
state radiation. This implies that in this formalism the initial chains
generally contain a larger number of links, and that final state radiation
is correspondingly allowed in a more restricted kinematic region. A 
consequence is that a larger set of chains contribute in the calculation of 
$\mathcal{G}$, with each 
chain given a smaller weight. This reduction is represented by a ``non-eikonal
form factor'', in such a way that the BFKL result is reproduced to leading
log accuracy. For further discussions of this difference between the formalisms
see refs. \cite{LDC} and \cite{salam}.

\section{Linked Dipole Chain model}

\subsection{LDC for DIS}

The Linked Dipole Chain (LDC) model
\cite{LDC} is a reformulation and
generalization of the CCFM model. Thus it is based on perturbative QCD 
calculations, and can be regarded as a formal quantum mechanical derivation 
of the semiclassical results presented above.

The LDC model is based on the observation that the
dominant features of the parton evolution is determined by a subset of
emitted gluons, which are ordered in both positive and negative
light-cone components, and also satisfy the relation
\begin{equation}
q_{\perp,i} > \min(k_{\perp,i},k_{\perp,i-1}).
\end{equation}
\label{eq:ldccut1}
In LDC this subset of ``\emph{primary}'' gluons
forms a chain of initial state radiation, and all other emissions are 
treated as final state radiation (including those emissions, which would
be emitted from a virtual link $k_i$, rather than from a quasireal
emission $q_i$, as discussed in section 5.5).
In ref.~\cite{LDC} it is shown that
adding the contributions from all different CCFM-chains with the same primary 
gluons, with their non-eikonal form factors, gives the following simple result:
\begin{equation}
\mathcal{G} \approx \sum_n \prod_i^n \int\!\!\int \bar{\alpha} \frac{d z_i}{z_i}
\frac{d^2 q_{\perp,i}}{\pi q_{\perp,i}^2} 
\theta(q_{\perp,i}-\min(k_{\perp,i},k_{\perp,i-1}))
\delta(x-\prod^n z_j)
\label{LDCekv}
\end{equation}
where the link momenta $\mathbf{k}_{\perp,i} = \sum_j^i \mathbf{q}_{\perp,j}$
are determined by transverse momentum conservation.
This implies that
\begin{equation}
q_{\perp,i}^2 \approx \max(k_{\perp,i}^2,k_{\perp,i-1}^2)
\label{LDCqtcut}
\end{equation}
From this relation we find for a step up or down in $k_\perp$:
\begin{itemize}
\item
Step up in $k_\perp$: $k_{\perp,i}>k_{\perp,i-1} \Rightarrow 
q_{\perp,i}\approx k_{\perp,i}$, which implies
\begin{equation}
\frac{d^2 q_{\perp,i}}{q_{\perp,i}^2} \approx
\frac{d^2 k_{\perp,i}}{k_{\perp,i}^2}
\label{stepup}
\end{equation}

\item
Step down in $k_\perp$: $k_{\perp,i}<k_{\perp,i-1} \Rightarrow 
q_{\perp,i}\approx k_{\perp,i-1}$, which implies
\begin{equation}
\frac{d^2 q_{\perp,i}}{q_{\perp,i}^2} \approx
\frac{d^2 k_{\perp,i}}{k_{\perp,i-1}^2} =
\frac{d^2 k_{\perp,i}}{k_{\perp,i}^2} \cdot 
\frac{k_{\perp,i}^2}{k_{\perp,i-1}^2}
\label{stepdown}
\end{equation}
\end{itemize}

We here recognize the suppression factor $k_{\perp,i}^2/k_{\perp,i-1}^2$ 
in eq.~(\ref{dubbelslutstegner}), associated with a step down in $k_\perp$.

Non-leading contributions from $1/(1-z)$-poles and quark links are added with 
Sudakov form factors.
In this formalism it is natural to include a running coupling
\begin{equation}
\alpha_s(q_{\perp,i}^2)=\alpha_s (\max(k_{\perp,i}^2,k_{\perp,i-1}^2))
\label{alfas-arg}
\end{equation}

Note, however, that this increase of $\alpha_s$ for soft emissions 
necessarily implies that a cutoff is needed for soft $k_\perp$ 
\cite{softcutoff}. This 
dividing line between the perturbative and the nonperturbative regimes has to 
be adjusted by fits to experimental data.

\begin{figure}
  \begin{center}
    \epsfig{figure=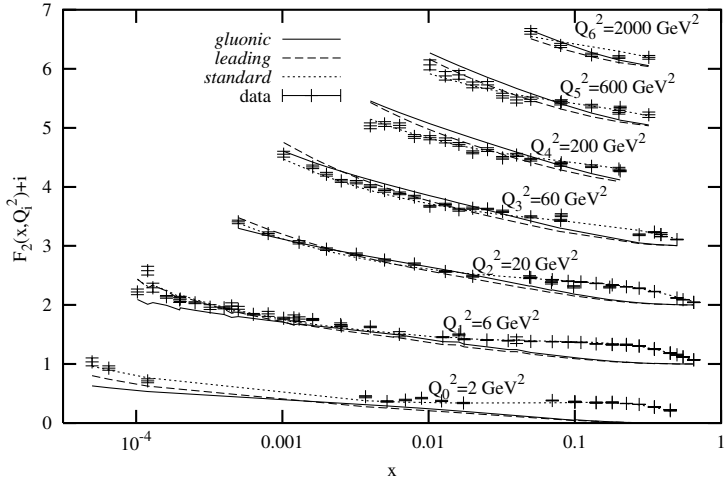,width=10cm}
  \end{center}
\caption{The description of $F_2$ data as a
  function of $x$ for different $Q^2$. 
   To separate the results, what is shown is
  $F_2+i$ for $Q_i^2$. Both small and large $x$ data from H1,
  ZEUS, NMC and E665 are included \cite{F2data}.  The favoured fit is the one 
  called standard and denoted by dotted lines. In this fit both gluon 
  and quark links are included. For a detailed description of the fits, 
  see ref.~\cite{gluondistrib}.}
\label{fig:f2}
\end{figure}

\begin{figure}
    \hskip -4mm \epsfig{file=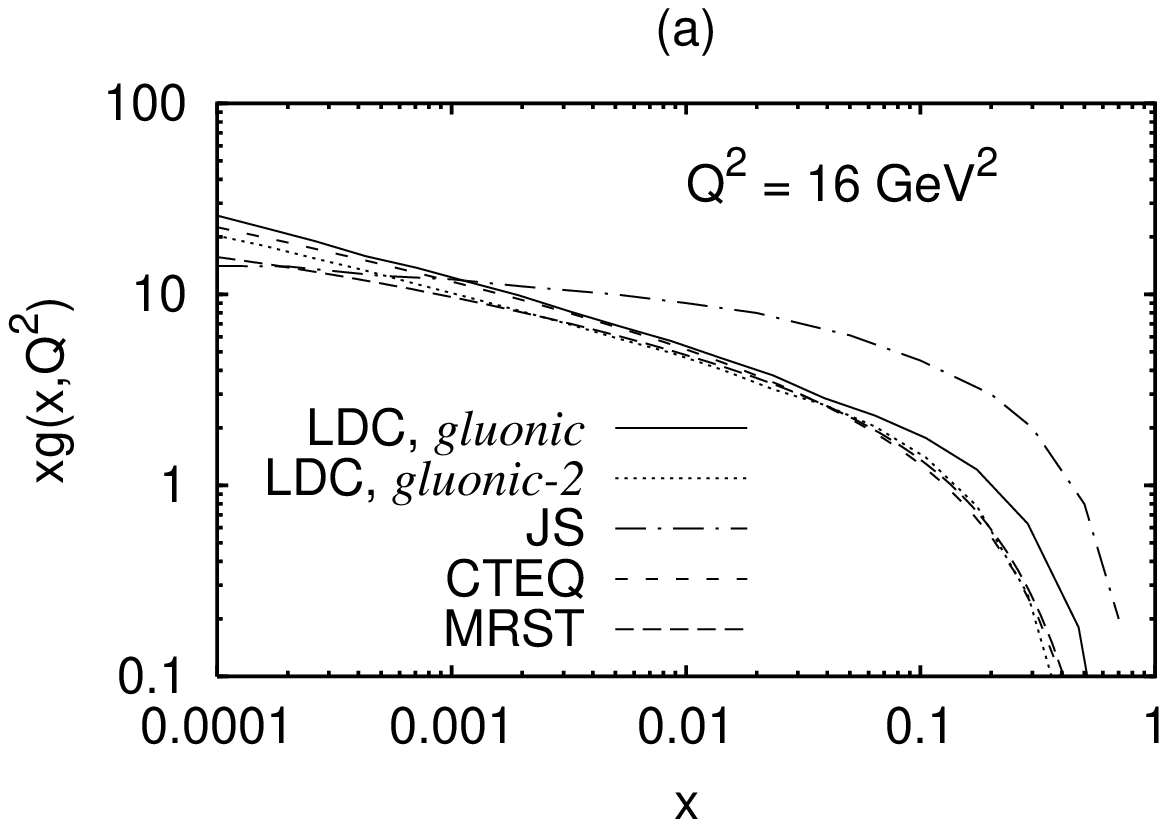,width=7cm}
    \hskip -6mm \epsfig{file=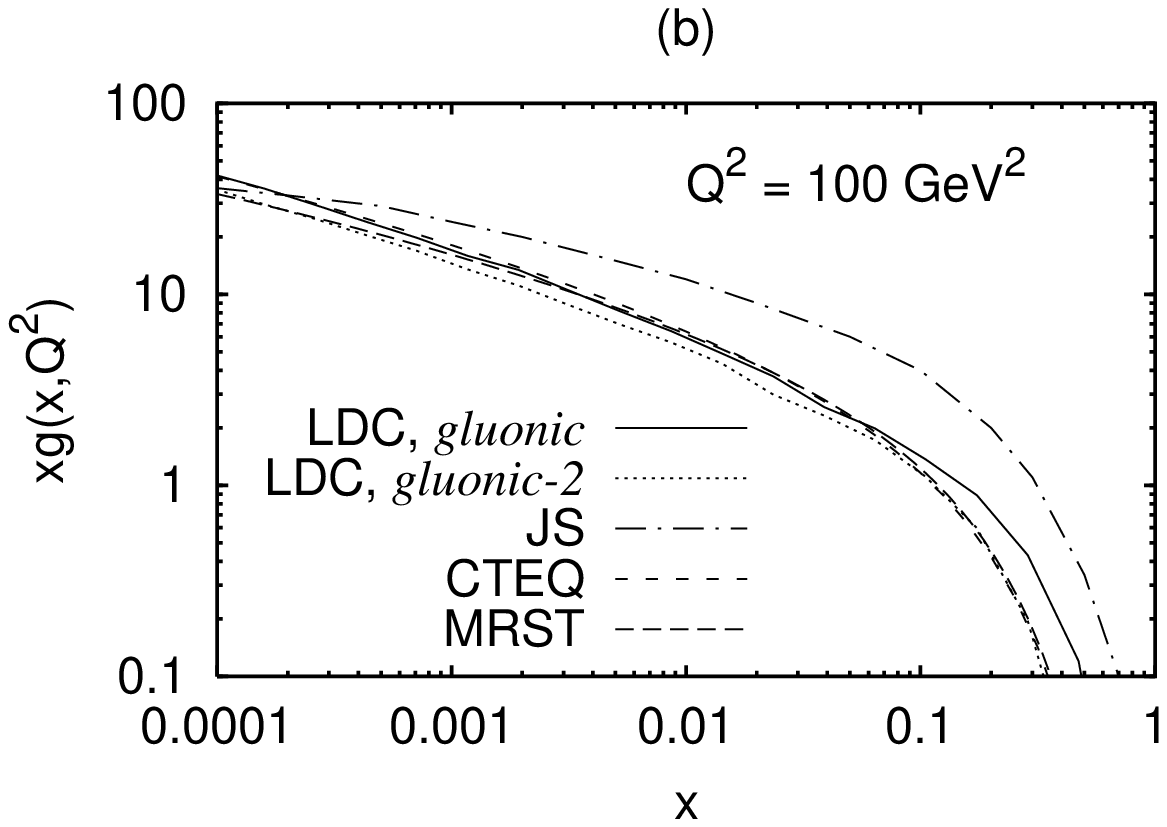,width=7cm}
\caption{The LDC integrated gluon
  distribution function for two different fits (called ``gluonic'' and 
  ``gluonic-2''), compared to the corresponding results from \textsc{Cascade}
  denoted JS (dash-dotted curve), CTEQ (short-dashed curve) and MRST
  (long-dashed curve), for (a) $Q^2 = 16$~GeV$^2$ and (b) $Q^2 =
  100$~GeV$^2$.}
\label{fig:int-comp}
\end{figure}

A Monte Carlo implementation, called \textsc{ldcmc}, is developed by H. Kharraziha and 
L. L\"{o}nnblad \cite{LDCMC}. This program does reproduce a large set of
experimental data. Fig.~\ref{fig:f2} shows a fit 
to $F_2$ \cite{gluondistrib}, compared with data from HERA and fixed target
experiments. The corresponding gluon distribution is shown in 
fig.~\ref{fig:int-comp},
and we note that this result shows good agreement with the results from the
CTEQ \cite{CTEQ} and MRST \cite{MRST} fits.
(Note that the LDC result is fitted 
to DIS data only, while the CTEQ and MRST fits include data from hadronic 
collisions.) As mentioned above, the LDC model is a generalization of
the CCFM model, which is implemented in the event generator \textsc{Cascade} 
\cite{cascade}. Both \textsc{ldcmc} and \textsc{Cascade} reproduce 
the main features of the final state properties \cite{forward}. 
A problem is, however, that
both models are able to reproduce the forward jet cross section only if the 
non-singular terms in the splitting functions are omitted \cite{forward}.
We do not know 
whether these terms are compensated by some dynamical mechanism, or if the
modelling of the proton fragmentation has to be improved.

\subsection{Hadronic collisions}

The LDC formalism can be applied also to hadron-hadron collisions
\cite{hadroncoll,mult}. In hadronic 
collisions \emph{multiple interactions} are 
needed to describe associated multiplicity and transverse energy in events 
with high $p_\perp$ jets. Such multiple interactions can originate either
from two hard 
subcollisions in a single parton chain, \emph{or} from more than one chain 
in a single event.

\begin{figure}
  \begin{center}
    \epsfig{figure=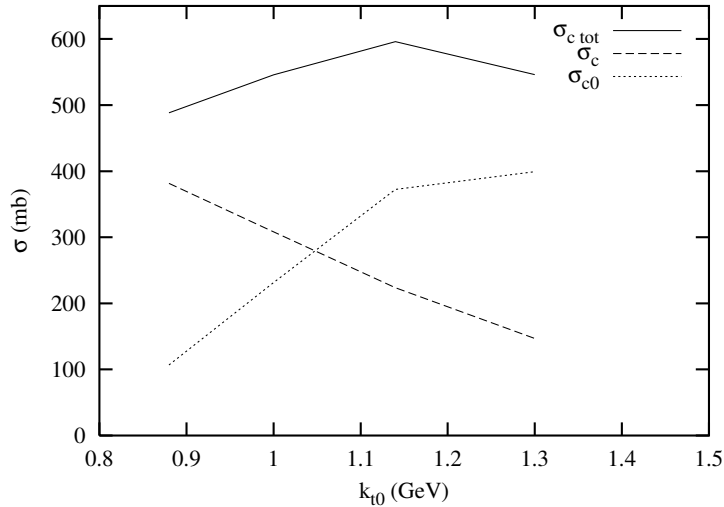,width=10cm}
  \end{center}
\caption{The cross section per chain in the LDC
  model as a function of the cutoff, $k_{\perp0}^2$. The dashed line
  is the cross section for chains with at least on emission above
  the cutoff, the dotted line is for soft chains without emissions
  above the cutoff, and the full line is the sum of the two. Note
  that the input parton densities have been re-fitted for each value
  of $k_{\perp0}$.}
\label{fig:sigkt0}
\end{figure}

The LDC formalism is very suitable for describing such events. A potential
problem arises because good fits to DIS data can be obtained for different 
values for the low $k_\perp$ cutoff, provided that the soft input structure 
function is adjusted accordingly. This is important because
the number of hard chains depends on this soft
cutoff. If the cutoff is increased a more singular input gluon distribution is 
needed in the fit to $F_2$. These soft input gluons, present at scale 
$Q_0^2$, may also form ``soft chains'' between the 
two colliding protons. These soft chains then have no emitted gluons with
$q_\perp$ above $Q_0$, but they do produce hadrons with low $p_\perp$. 
From the fits to DIS data we then find that
the reduction in the number of hard chains is just compensated by an increase
in the number of soft chains, as is illustrated in fig~\ref{fig:sigkt0}.
As a result we see that, in this formalism, the (average) total number of
chains in a $pp$ collision can be fixed by fits to DIS data. This result 
apparently implies a very strong connection between DIS and hadronic 
collisions \cite{mult}.

This formalism should also be applicable in studies of effects of unitarisation
and saturation. Separate branches of a chain can interact
with a colliding hadron, or different chains can
join each other, and these possibilities are presently under study. Also 
applications to diffractive collisions are of interest.

\section{Conclusions}

We have seen that the behaviour of DIS in the DGLAP and BFKL domains can be 
understood in a semiclassical intuitive picture. A more quantitative 
description is offered by the LDC formalism, which smoothly interpolates between
these two kinematical regions. 

The formalism also offers a link between DIS and hadronic collisions. Within 
the formalism fits to data on $F_2$ in DIS can be used to predict 
the number of chains and hard subcollisions in a high energy 
proton-proton collision. 

Further work on multiple interaction and saturation is in progress, as 
well as studies of diffractive scattering.

\section{Acknowledgement}

I want to thank my collaborators Leif L\"{o}nnblad and Gabriela Miu.

\end{document}